\documentclass[twocolumn,superscriptaddress,showkeys]{revtex4}
\usepackage{amssymb,epsf,epsfig,amsmath,color}
\usepackage{multirow}
\usepackage{accents}

\newcommand{\lsim}{
\mathrel{\hbox{\rlap{\hbox{\lower4pt\hbox{$\sim$}}}\hbox{$<$}}}}
\newcommand{\gsim}{
\mathrel{\hbox{\rlap{\hbox{\lower4pt\hbox{$\sim$}}}\hbox{$>$}}}}

\usepackage{xspace}

\newcommand{\B}{\ensuremath{B}\xspace}
\newcommand{\Bbar}{\kern 0.18em\overline{\kern -0.18em B}{}\xspace}

\newcommand{\Bd }{\ensuremath{\B^0_d}\xspace}

\newcommand{\Bs }{\ensuremath{\B^0_s}\xspace}

\newcommand{\uBs}{\ensuremath{\B_s^{\vphantom{+}}}\xspace}

\newcommand{\Kbar}{\kern 0.18em\overline{\kern -0.18em K}{}\xspace}

\newcommand{\Bdtopipi}{\mbox{\ensuremath{\Bd\to \pi^-\pi^+}}\xspace}

\newcommand{\BdtoKpi}{\mbox{\ensuremath{\Bd\to \pi^-K^+}}\xspace}

\newcommand{\BstoKK}{\mbox{\ensuremath{\Bs\to K^-K^+}}\xspace}

\newcommand{\uBstoKK}{\mbox{\ensuremath{\uBs\to K^-K^+}}\xspace}

\def\theabstract{The $U$-spin symmetry provides a powerful tool to extract the angle $\gamma$ of the 
Unitarity Triangle and the $B_s^0$--$\bar{B}_s^0$ mixing phase $\phi_s$ from CP violation in
the $B^0_s\to K^-K^+$, $B^0_d\to\pi^-\pi^+$ system. LHCb has obtained first results with 
uncertainties at the $7^\circ$ level. Due to $U$-spin-breaking corrections, it will be challenging to 
reduce the uncertainty below ${\cal O}(5^\circ)$ at Belle II and the LHCb upgrade. We propose a 
new strategy, using $\gamma$ as input and utilizing $B^0_s\to K^-\ell^+\nu_\ell$, $B^0_d\to \pi^-\ell^+\nu_\ell$ 
decays, which allows an extraction of $\phi_s$ with a future theoretical precision of up to ${\cal O}(0.5^\circ)$,
thereby matching the experimental prospects. Since $B^0_s\to K^-K^+$ is dominated by penguin topologies, 
new sources of CP violation may be revealed.}

\begin{document}
\begin{titlepage}
\vspace*{-0.1truecm}
\begin{flushright}
Nikhef-2016-038\\
QFET-2016-13\\
SI-HEP-2016-23
\end{flushright}

\vspace{1.3truecm}

\begin{center}
\boldmath
{\Large{\bf New Strategy to Explore CP Violation with $B^0_s\to K^-K^+$}}
\unboldmath
\end{center}

\vspace{1.2truecm}

\begin{center}
{\bf Robert Fleischer,\,${}^{a,b}$  Ruben Jaarsma,\,${}^{a}$ and  K. Keri Vos\,${}^{a,c,d}$}

\vspace{0.5truecm}

${}^a${\sl Nikhef, Science Park 105, NL-1098 XG Amsterdam, Netherlands}

${}^b${\sl  Department of Physics and Astronomy, Vrije Universiteit Amsterdam,\\
NL-1081 HV Amsterdam, Netherlands}

${}^c${\sl Van Swinderen Institute for Particle Physics and Gravity, University of Groningen,\\
NL-9747 AG Groningen, Netherlands}

${}^d${\sl Theoretische Physik 1, Naturwissenschaftlich-Technische Fakult\"at, \\
Universit\"at Siegen, D-57068 Siegen, Germany}

\end{center}

\vspace*{1.7cm}

\begin{center}
\large{\bf Abstract}\\

\vspace*{0.6truecm}

\begin{tabular}{p{13.5truecm}}
\theabstract
\end{tabular}

\end{center}

\vspace*{2.1truecm}

\vfill

\noindent
August 2016

\end{titlepage}

\newpage
\thispagestyle{empty}
\mbox{}

\newpage
\thispagestyle{empty}
\mbox{}

\rule{0cm}{23cm}

\newpage
\thispagestyle{empty}
\mbox{}

\setcounter{page}{0}

\preprint{Nikhef-2012-nnn}

\date{\today}

\title{\boldmath New Strategy to Explore CP Violation with $B^0_s\to K^-K^+$
\unboldmath}

\author{Robert Fleischer}
%\email{Robert.Fleischer@nikhef.nl}
\affiliation{Nikhef, Science Park 105, NL-1098 XG Amsterdam, Netherlands}
\affiliation{Department of Physics and Astronomy, Vrije Universiteit Amsterdam,
NL-1081 HV Amsterdam, Netherlands}

\author{Ruben Jaarsma}
%\email{Ruben.Jaarsma@nikhef.nl}
\affiliation{Nikhef, Science Park 105, NL-1098 XG Amsterdam, Netherlands}

\author{K. Keri Vos}
%\email{Keri.Vos@uni-siegen.de}
\affiliation{Nikhef, Science Park 105, NL-1098 XG Amsterdam, Netherlands}
\affiliation{Van Swinderen Institute for Particle Physics and Gravity, University of Groningen,\\
NL-9747 AG Groningen, Netherlands}
\affiliation{Theoretische Physik 1, Naturwissenschaftlich-Technische Fakult\"at, Universit\"at Siegen, 
D-57068 Siegen, Germany}

	\date{\today}
	\vspace{3em}

\begin{abstract}
\vspace{0.2cm}\noindent
\theabstract 
\end{abstract}


\maketitle

\thispagestyle{empty}
%	\vbox{}
%
%
%	
\setcounter{page}{1}
\section{Introduction} 
Decays of $B$ mesons offer an interesting laboratory to search for signals of physics beyond
the Standard Model (SM). In particular the penguin sector is sensitive to new heavy particles, 
which may enter the corresponding loop diagrams or cause flavor-changing neutral currents 
at the tree level \cite{BuGi}. Such new interactions are usually associated with new sources of CP violation,
which would manifest themselves in CP-violating decay rate asymmetries. 

These CP asymmetries are induced through interference effects. 
Interference between different decay contributions, such as tree and penguin
topologies, results in direct CP violation. In the case of neutral $B^0_q$ mesons ($q=d,s$), interference 
between $B^0_q\to f$ and $\bar B^0_q\to f$ transitions through $B^0_q$--$\bar B^0_q$ mixing may
generate mixing-induced CP violation \cite{RF-rev}. In order to detect footprints of New Physics (NP) 
in the era of Belle II \cite{Belle-II} and the LHCb upgrade \cite{LHCb-strategy}, the SM picture of the CP 
asymmetries has to be understood with highest precision, where the main challenge is related to 
the impact of strong interactions. 

The $B^0_s\to K^-K^+$ mode is one of the most prominent non-leptonic $B$ decays, receiving contributions 
from tree and penguin topologies. Due to the specific pattern of the quark-flavor mixing in the SM, 
which is encoded in the Cabibbo--Kobayashi--Maskawa (CKM) matrix \cite{RF-rev}, the latter loop 
processes play the dominant role.

The $B^0_s\to K^-K^+$ channel is related to $B^0_d\to\pi^-\pi^+$ through the $U$-spin flavor 
symmetry of strong interactions, relating down and strange quarks to each other. Exploiting this 
feature, the hadronic non-perturbative parameters of $B^0_s\to K^-K^+$, which suffer from large 
theoretical uncertainties, can be related to their counterparts in $B^0_d\to\pi^-\pi^+$, 
allowing the extraction of the angle $\gamma$ of the Unitarity Triangle and the $B^0_s$--$\bar B^0_s$ 
mixing phase $\phi_s$ \cite{RF-99,RF-07,FK}.  A variant of this $U$-spin method was proposed in 
\cite{CFMS}, combining it with the $B\to \pi\pi$ isospin analysis \cite{GL}, which reduces the sensitivity to 
$U$-spin-breaking effects. 

Using their first measurement of CP violation in $B^0_s\to K^-K^+$  \cite{LHCb-BsKK-CP}, the 
LHCb collaboration \cite{LHCb-BsKK-gam} obtained
\begin{equation}\label{LHCb-res}
\gamma=(63.5^{+7.2}_{-6.7})^\circ, \quad \phi_s=-(6.9^{+9.2}_{-8.0})^\circ.
\end{equation}
In this analysis, the strategies proposed in \cite{RF-99} and \cite{CFMS} were found to agree 
with each other and previous studies  \cite{RF-07,FK} for $U$-spin-breaking effects of up to 50\%. 
For even larger corrections of (50--100)\%, the $B\to\pi\pi$ system stabilizes the situation. 

Using pure tree decays $B\to D^{(*)} K^{(*)}$, $\gamma$ can be extracted in a theoretically clean way 
\cite{FR-gam}. Current data yield the averages $\gamma=(73.2_{-7.0}^{+6.3})^{\circ}$ \cite{CKMfitter}
and $(68.3\pm7.5)^{\circ}$ \cite{UTfit}, which agree with (\ref{LHCb-res}) and have similar 
uncertainties. The phase $\phi_s$ takes the SM value $\phi_s^{\rm SM}=-(2.1\pm0.1)^\circ$ 
\cite{ABL} and can be determined through $B^0_s\to J/\psi \phi$ and similar decays which are 
dominated by tree topologies; penguin contributions limit the theoretical precision 
(see \cite{DeBF} and references therein). The Particle Data Group (PDG) gives the average 
$\phi_s=-(0.68\pm2.2)^\circ$ \cite{PDG}, which has an uncertainty about four times smaller than 
(\ref{LHCb-res}). In the future, the uncertainty for $\gamma$ from tree decays can be reduced to 
${\cal O}(1^\circ)$ \cite{Belle-II,LHCb-strategy}, while $\phi_s$ can be determined from 
$B^0_s\to J/\psi \phi$ and penguin control channels with a precision at the $0.5^\circ$ level \cite{DeBF}. 

The experimental results in (\ref{LHCb-res}) suggest significant room for improvement. However, 
the theoretical precision is limited by $U$-spin-breaking corrections to penguin topologies.
As we will show, it is challenging to reduce the uncertainty below ${\cal O}(5^\circ)$. 
We propose a new strategy to exploit the physics potential of $B^0_s\to K^-K^+$, $B^0_d\to\pi^-\pi^+$  
in the high-precision era of $B$ physics. It employs semileptonic $B^0_s\to K^-\ell^+\nu_\ell$, 
$B^0_d\to \pi^-\ell^+\nu_\ell$ decays as new ingredients and applies the $U$-spin symmetry only
to theoretically well-behaved quantities. This method will eventually allow a measurement of $\phi_s$ 
from CP violation in $B^0_s\to K^-K^+$ with a theoretical precision at the $0.5^\circ$ level, thereby 
matching the expected experimental precision. It has the exciting potential to reveal CP-violating NP 
contributions to the penguin-dominated $B^0_s\to K^-K^+$ mode, and provides valuable insights into 
strong interaction dynamics through the determination of $U$-spin-breaking parameters.

\begin{table}
	\centering
	\begin{tabular}{l | r | r }
		Observable & Current \cite{PDG, HFAG}  & LHCb upgrade \cite{LHCb-strategy} \\
		\hline\hline
		$\mathcal{A}_{\textrm{CP}}^{\textrm{dir}}(B_d\rightarrow \pi^-\pi^+)$ & $ -0.31 \pm 0.05$ & $-0.3 \pm 0.008$\\
		$\mathcal{A}_{\textrm{CP}}^{\textrm{mix}}(B_d\rightarrow \pi^-\pi^+)$ & $0.66 \pm 0.06 $ & $0.66 \pm 0.008$ \\
				$\mathcal{A}_{\textrm{CP}}^{\textrm{dir}}(B_s\rightarrow K^-K^+)$ & $0.14 \pm 0.11$ & $0.085 \pm 0.008$\\
		$\mathcal{A}_{\textrm{CP}}^{\textrm{mix}}(B_s\rightarrow K^-K^+)$ & $-0.3\pm 0.13$ & $-0.19 \pm 0.008$ 	
	\end{tabular}
	\caption{Summary of the current and future measurements. For the upgrade scenario, we use
	$(d,\theta)$ following from the CP asymmetries of $B^0_d\to \pi^-\pi^+$ to calculate the central
	values of the  $B^0_s\to K^-K^+$ CP asymmetries with the $U$-spin symmetry.}
	\label{table}
\end{table}

\section{The Original Strategy}\label{sec:class}
Before focusing on the new method, it is instructive to have a look at the original strategy
\cite{RF-99,RF-07,FK}. In the SM, the $B^0_s\to K^-K^+$ decay amplitude can be written as 
\begin{equation}
A\left(\BstoKK\right) = e^{i\gamma}\sqrt{\epsilon}\,\mathcal{C}'
\left[1+\frac{1}{\epsilon}d' e^{i\theta'}e^{-i\gamma}\right],
\end{equation}
where the primes indicate a $\bar b \to \bar s$ transition, and
\begin{equation}\label{Cprime}
\mathcal{C}' = \lambda^3A\,R_b\left[T' + E' + P^{(ut)'} + PA^{(ut)'}\right]
\end{equation}
\begin{equation}\label{dprime}
d'e^{i\theta'} \equiv \frac{1}{R_b}\left[\frac{P^{(ct)'}+PA^{(ct)'}}{T'+E'+P^{(ut)'}+PA^{(ut)'}}\right]
\end{equation}
with
\begin{equation}
P^{(qt)'}\equiv P^{(q)'}-P^{(t)'}, \quad PA^{(qt)'}\equiv PA^{(q)'}-PA^{(t)'}.
\end{equation}
Here $T'$ is a colour-allowed tree and $E'$ an exchange amplitude, while $P^{(q)'}$ and 
$PA^{(q)'}$ denote penguin and penguin annihilation topologies, respectively, 
with $q=u,c,t$ quarks in the loops. Finally, $A\equiv|V_{cb}|/\lambda^2\approx0.8$, 
$R_b \equiv \left(1-\lambda^2/2\right)|V_{ub}/(\lambda V_{cb})|\approx 0.4$ and
$\epsilon \equiv \lambda^2/(1-\lambda^2) \approx 0.05$ are CKM factors 
involving the Wolfenstein parameter $\lambda \equiv |V_{us}|\approx 0.22$ \cite{CKMfitter}. 
The exchange and penguin annihilation topologies are expected to play a minor role on the basis of
dynamical arguments \cite{GHLR,GHLR-SU3,BGV}. 

The amplitude of the $\bar b\to \bar d$ mode $B^0_d\to\pi^-\pi^+$ reads
\begin{equation}
A\left(\Bdtopipi\right) = e^{i\gamma}\mathcal{C}\left[1- d e^{i\theta}e^{-i\gamma}\right],
\end{equation}
where the hadronic parameters $\mathcal{C}$ and $d e^{i\theta}$ are defined in analogy to their 
$B^0_s\to K^-K^+$ counterparts. 
The $U$-spin symmetry implies the following relations \cite{RF-99}:
\begin{align}
d'e^{i\theta'} &= d e^{i\theta} \label{d-rel} \\
\mathcal{C}'& =\mathcal{C} \label{C-rel}.
\end{align}

Due to $B^0_q$--$\bar B^0_q$ oscillations, we obtain time-dependent decay rate asymmetries which 
probe direct and mixing-induced CP violation, described by ${\cal A}_{\rm CP}^{\rm dir}(B_q\to f)$ and 
${\cal A}_{\rm CP}^{\rm mix}(B_q\to f)$, respectively \cite{RF-rev}. In the case of $B^0_s\to K^-K^+$ 
and $B^0_d\to\pi^-\pi^+$, these observables depend -- in 
addition to $\gamma$ -- on $d'e^{i\theta'}$ and $d e^{i\theta}$; the ${\cal A}_{\rm CP}^{\rm mix}(B_q\to f)$
involve also the $B^0_q$--$\bar B^0_q$ mixing phases $\phi_q$. 

The main application of this system is usually the determination of $\gamma$, using the $\phi_q$ as input. 
However, if only $\phi_d$ is employed, also $\phi_s$ can be extracted. 
In view of the large uncertainties of the current LHCb measurement of the $B^0_s\to K^-K^+$ CP 
asymmetries (see Table~\ref{table}), the results in (\ref{LHCb-res}) are governed by the CP asymmetries 
of $B^0_d\to\pi^-\pi^+$ and the ratio of the branching ratios of $B^0_s\to K^-K^+$ and $B^0_d\to\pi^-\pi^+$ 
\cite{RF-07,FK}. The latter is affected by $U$-spin-breaking corrections to (\ref{C-rel}) which involve 
non-perturbative decay constants and form factors.

\begin{figure}[t]
		\centering
	\includegraphics[width = 0.8\linewidth]{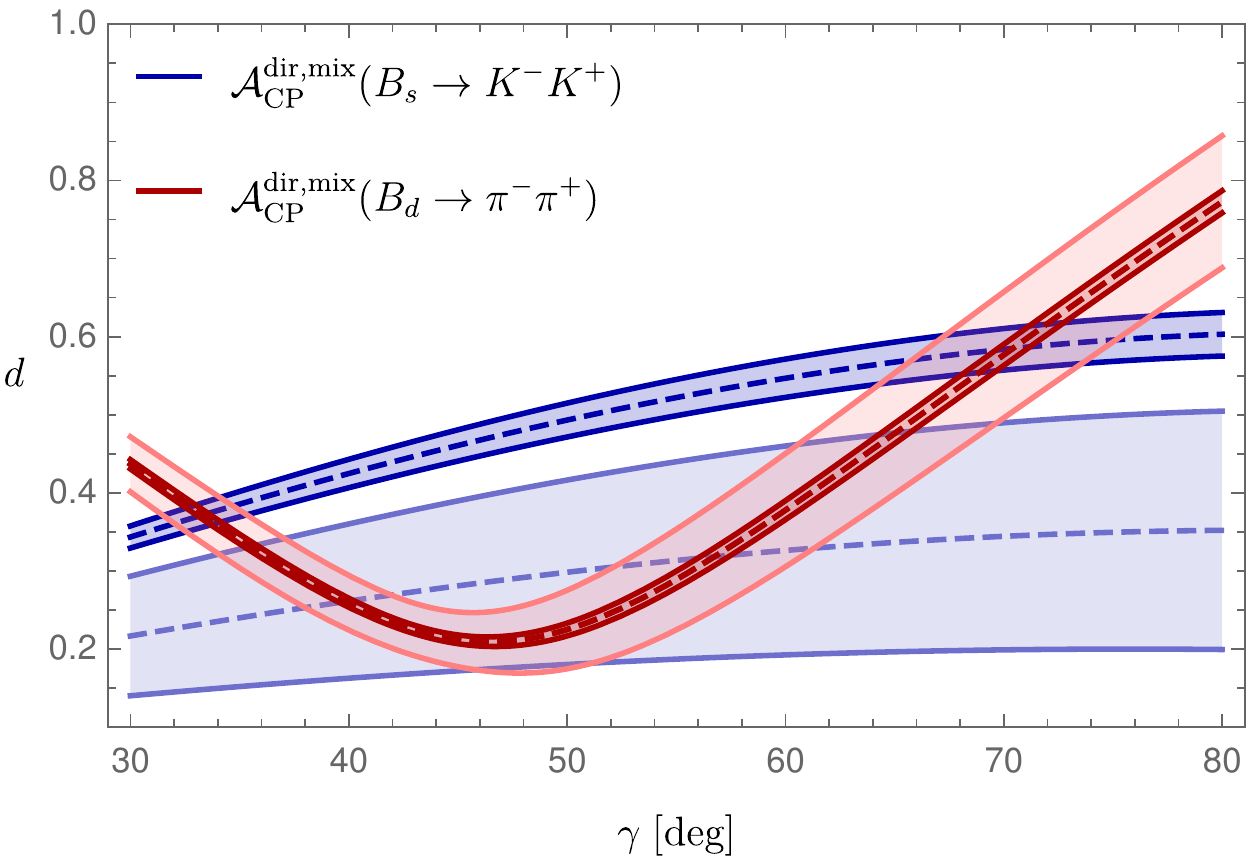}
	\caption{Illustration of the $\gamma$ determination from the CP asymmetries of the
	$B^0_d\to\pi^-\pi^+$, $B^0_s\to K^-K^+$ decays: current situation (wide bands), LHCb upgrade 
	(narrow bands).}\label{fig:d-theta-classic}
\end{figure}

At the LHCb upgrade, $\gamma$ can be extracted by using only the CP asymmetries. In this
case, the $U$-spin relation $d'=d$ is sufficient, which is more favorable than (\ref{C-rel}) because
factorizable $U$-spin-breaking corrections cancel \cite{RF-99}. In Table~\ref{table}, we collect 
expected measurements of the 
CP asymmetries at the LHCb upgrade.  Assuming the $U$-spin relation $d'=d$, 
we get an experimental uncertainty of $\gamma$ of ${\cal O}(1^\circ)$, as illustrated in 
Fig.~\ref{fig:d-theta-classic}. Allowing for $U$-spin-breaking 
corrections as
\begin{equation}\label{U-break}
\xi\equiv \frac{d'}{d} = 1\pm 0.2, \quad \Delta\equiv \theta'-\theta = (0\pm20)^\circ,
\end{equation}
where only $\xi$ affects the determination of $\gamma$, gives an uncertainty of
${\cal O}(5^\circ)$.

The CP asymmetries of $B^0_s\to K^-K^+$ allow the determination of the following ``effective" 
mixing phase \cite{FK-DG}:
\begin{equation}\label{phis-eff}
\sin{\phi_s^{\textrm{eff}}}
=\frac{\mathcal{A}_{\rm CP}^{\rm mix}(\uBstoKK)}{\sqrt{1-\mathcal{A}_{\rm CP}^{\rm dir}(\uBstoKK)^2}} \ ,
\end{equation}
where $\phi_s^{\textrm{eff}}\equiv\phi_s+\Delta \phi_{KK}$ with 
\begin{equation}\label{DelPhiKK}
\hspace*{-0.1truecm}
\tan \Delta \phi_{KK} = \epsilon\left[\frac{2(d'\cos\theta'+\epsilon\cos\gamma)\sin\gamma}{d'^2+
2\epsilon d'\cos\theta'\cos\gamma+\epsilon^2\cos2\gamma}\right].
\end{equation}
At the LHCb upgrade, $\phi_s^{\textrm{eff}}$ can be measured with a precision at the $0.5^\circ$ level
\cite{LHCb-strategy}. Using $\gamma=(70\pm1)^\circ$ and the CP asymmetries of 
$B^0_d\to\pi^-\pi^+$ as input, $d$ and $\theta$ can be extracted. If we assume $U$-spin-breaking 
corrections as given by (\ref{U-break}) when converting the $B^0_d\to\pi^-\pi^+$ parameters into their 
$B^0_s\to K^-K^+$ counterparts, we obtain an uncertainty for $\Delta \phi_{KK}$ of $2.6^\circ$, 
which affects the determination of $\phi_s=\phi_s^{\textrm{eff}}-\Delta \phi_{KK}$ 
correspondingly. 

In order to match the future experimental precision, $\xi$ would have to be known with an uncertainty at the 
few percent level. Unless there is theoretical progress, this precision is out of reach and the 
impressive experimental prospects at the LHCb upgrade cannot be fully exploited.

\section{The New Strategy}
Our goal is to make minimal use of the $U$-spin symmetry. We employ $\gamma$ as an input, assuming 
$\gamma=(70\pm1)^\circ$ as determined from pure tree decays in the era of Belle II and the LHCb upgrade 
\cite{Belle-II,LHCb-strategy}. Moreover, we use $\phi_d$ as an input, which can be extracted from 
$B^0_{d,s}\to J/\psi K_{\rm S}$ decays taking penguin contributions into account, assuming 
$\phi_d=(43.2\pm0.6)^\circ$ \cite{DeBF}. The CP asymmetries of $B^0_d\to\pi^-\pi^+$ allow then 
a theoretically clean determination of the hadronic parameters $d$, $\theta$ and ${\cal C}$. We shall
focus on the determination of $\phi_s$ from (\ref{phis-eff}) which requires knowledge of the 
hadronic phase shift $\Delta\phi_{KK}$ in (\ref{DelPhiKK}). 

The ratios of non-leptonic decay rates to differential semileptonic rates allow us to probe non-factorizable 
effects of strong interactions \cite{BJ,BS,rosner,NS,BBNS,FST-fact}. In analogy to the analysis of
$B\to D\bar D$ decays in \cite{BDD}, we introduce
\begin{displaymath}
R_{\pi} \equiv 
\frac{ \Gamma(B_d\rightarrow \pi^-\pi^+)}{d\Gamma(B^0_d\rightarrow \pi^- \ell^+ \nu_\ell)
/dq^2|_{q^2=m_\pi^2}} 
\end{displaymath}
\begin{equation}\label{Rpi}
=6\pi^2|V_{ud}|^2f_\pi^2X_\pi \, r_\pi \,|a_{\rm NF}|^2,
\end{equation}
where $|V_{ud}|$ is a CKM matrix element, $f_\pi$ denotes the charged pion decay constant, 
\begin{equation}\label{Xpi}
X_\pi= \left[\frac{(m_{B_d}^2-m_{\pi}^2)^2}{m_{B_d}^2(m_{B_d}^2-4 \, m_{\pi}^2)}\right] 
\left[\frac{F_0^{B_d\pi}(m_{\pi}^2)}{F_1^{B_d\pi}(m_{\pi}^2)} \right]^2
\end{equation}
depends on the meson masses and form factors, 
\begin{equation}
r_\pi = 1-2d\cos\theta\cos\gamma+d^2,
\end{equation}
and 
\begin{equation}
a_{\rm NF}=a_{\rm NF}^T (1+r_P)(1+x)
\end{equation}
characterizes non-factorizable effects with
\begin{equation}\label{rP-x-def}
r_P\equiv \frac{P^{(ut)}}{T}, \quad x\equiv\frac{E+PA^{(ut)}}{T+P^{(ut)}}.
\end{equation}
The deviation of $a_{\rm NF}^T $ from one characterizes non-factorizable contributions to $T$.  
From the theoretical point of view, this color-allowed tree amplitude is most favorable, while the 
penguin topologies are challenging, with issues such as ``charming penguins" \cite{charm-pen}. 
The framework of QCD factorization sets a stage for the theoretical description \cite{QCDF,BeNe}, 
where two-loop next-to-next-to-leading-order vertex corrections were calculated \cite{BHL}:
\begin{equation}\label{QCDF-calc}
a_{\rm NF}^T=1.000^{+0.029}_{-0.069}+(0.011^{+0.023}_{-0.050})i.
\end{equation}

In analogy to (\ref{Rpi}), we introduce
\begin{displaymath}
{R}_K \equiv \frac{\Gamma(B_s \to K^-K^+)}{d\Gamma(B_s^0 \to K^-\ell^+\nu_\ell)/dq^2|_{q^2=m_K^2}} 
\end{displaymath}
\begin{equation}\label{RK}
=6\pi^2|V_{us}|^2f_K^2X_K \, r_K \, |a_{\rm NF}'|^2,
\end{equation}
where
\begin{equation}\label{rK-def}
r_K = 1+2\left(\frac{d'}{\epsilon}\right)\cos\theta'\cos\gamma+\left(\frac{d'}{\epsilon}\right)^2 ,
\end{equation}
and $X_K$ can be obtained from (\ref{Xpi}) through straightforward replacements.

\begin{figure}[t]
		\centering
	\includegraphics[width = 0.9\linewidth]{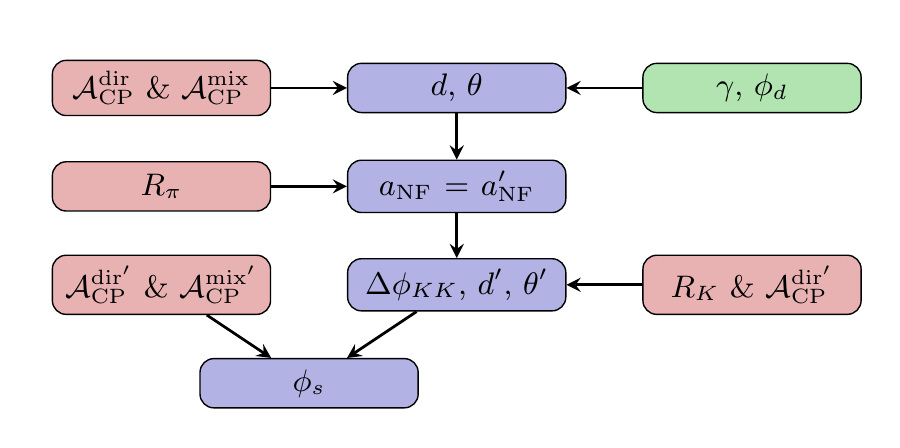}
	\caption{Illustration of the new strategy to extract $\phi_s$ from CP violation 
	in $\BstoKK$. The $\mathcal{A}^{\rm dir}_{\rm CP}$, $\mathcal{A}^{\rm mix}_{\rm CP}$ and 
	$\mathcal{A}^{\rm dir '}_{\rm CP}$, $\mathcal{A}^{\rm mix '}_{\rm CP}$ denote the direct, 
	mixing-induced CP asymmetries of the $B^0_d\to\pi^-\pi^+$ and $B^0_s\to K^-K^+$ decays,
	respectively.}\label{fig:flowchart}
\end{figure}

From the CP asymmetries of $B^0_d\to\pi^-\pi^+$  we may determine $r_\pi$. The ratio of $R_K$ 
and $R_\pi$ yields then
\begin{equation}
r_K=r_\pi \frac{R_K}{R_\pi} \left(\frac{|V_{ud}|\,f_\pi}{|V_{us}|f_K}\right)^2\frac{X_\pi}{X_K}\left| 
\frac{a_{\rm NF}}{a_{\rm NF}'} \right|^2,
\end{equation}
which allows us to determine the observable in (\ref{rK-def}) by applying the $U$-spin symmetry to 
the following ratio: 
\begin{equation}\label{a-rel}
\xi_{\rm NF}^a\equiv\left|\frac{a_{\rm NF}}{a_{\rm NF}'}\right|=\left|\frac{a_{\rm NF}^T}{a_{\rm NF}^{T'}}\right|
\left|\frac{1+r_P}{1+r_P'}\right|\left|\frac{1+x}{1+x'}\right|.
\end{equation}
Experimental data for charged kaon and pion leptonic decays allow the determination of 
$|V_{us}| f_K/|V_{ud}| f_\pi = 0.27599\pm0.00037$ with impressive precision \cite{Ros15}. The double
ratio of form factors in $X_\pi/X_K$ is given with excellent precision by one, which is also in agreement
with the kinematic constraint implemented by lattice calculations \cite{lattice-1,lattice-2}. 
Using $\mathcal{A}_{\rm CP}^{\rm dir}(\uBstoKK)\equiv{\cal A}_{\rm CP}^{\rm dir'}$, 
which depends on $d'$, $\theta'$ and $\gamma$ \cite{RF-99}, we may determine $d'$ and $\theta'$:
\begin{displaymath}
d'=\epsilon\Biggl[r_K+\cos2\gamma \pm \Bigl[(r_K+\cos 2\gamma)^2
\end{displaymath}
\vspace*{-0.6truecm}
\begin{equation}
\hspace*{1.5truecm}-(r_K-1)^2-(r_K {\cal A}_{\rm CP}^{\rm dir'}/\tan\gamma)^2  \Bigr]^{1/2} \Biggr]^{1/2}   
\end{equation}
\vspace*{-0.2truecm}
\begin{equation}
\sin\theta'=\frac{\epsilon\, r_K{\cal A}_{\rm CP}^{\rm dir'}}{2 d'\sin\gamma}, \quad
\cos\theta'=\frac{\epsilon^2(r_K-1)-d'^2}{2\epsilon d'\cos\gamma}\,.
\end{equation}
Finally, we determine $\Delta\phi_{KK}$ through (\ref{DelPhiKK}), 
which allows the extraction of $\phi_s=\phi_s^{\rm eff}-\Delta\phi_{KK}$ from the CP 
asymmetries of $B^0_s\to K^+K^-$ entering (\ref{phis-eff}). This method is illustrated in the 
flowchart in Fig.~\ref{fig:flowchart}.

The theoretical precision of this method is limited by the $U$-spin-breaking corrections to (\ref{a-rel}).
Writing $a_{\rm NF}^{T(')}=1+\Delta_{\rm NF}^{T(')}$ with 
$\Delta_{\rm NF}^{T'}=\Delta_{\rm NF}^T(1- \xi_{\rm NF}^T)$ yields
\begin{equation}\label{aT-SU3}
\frac{a^T_{\rm NF}}{a_{\rm NF}^{T'}}=1+\Delta_{\rm NF}^{T}\xi_{\rm NF}^T+
{\cal O}((\Delta_{\rm NF}^{T})^2).
\end{equation}
The numerical value in (\ref{QCDF-calc}) corresponds to $\Delta_{\rm NF}^T\sim0.05$. Consequently, 
$\xi_{\rm NF}\sim0.2$, i.e.\ $U$-spin-breaking corrections of $20\%$, corresponds to a correction 
at the $1\%$ level to (\ref{aT-SU3}). 
In the case of $r_P$ defined in (\ref{rP-x-def}), 
we write in analogy $r_P'=r_P(1-\xi_{r_P})$, which gives
\begin{equation}\label{rP-SU3}
\frac{1+r_P}{1+r_P'} = 1+r_P\xi_{r_P} + {\cal O}(r_P^2).
\end{equation}
Using data, we expect $r_P\sim 0.3$, which agrees with general expectations 
\cite{GHLR,GHLR-SU3}. Assuming $\xi_{r_P}\sim0.2$ yields a correction at the $5\%$ level. A 
similar structure arises for the $U$-spin-breaking ratio of $(1+x)/(1+x')$, which involves the 
exchange and penguin annihilation amplitudes (\ref{rP-x-def}). These topologies are expected 
to play a minor role. Experimental data for $B^0_d\to K^+K^-$ and $B^0_s\to\pi^+\pi^-$ decays 
allow us to constrain these contributions \cite{RF-07,BIG}. Making assumptions for $x$, $x'$ in 
analogy to the discussion for $r_P$, $r_P'$ would give a correction at the $5\%$ level. 

\begin{figure}[t]\centering
	\includegraphics[width = 0.9 \linewidth]{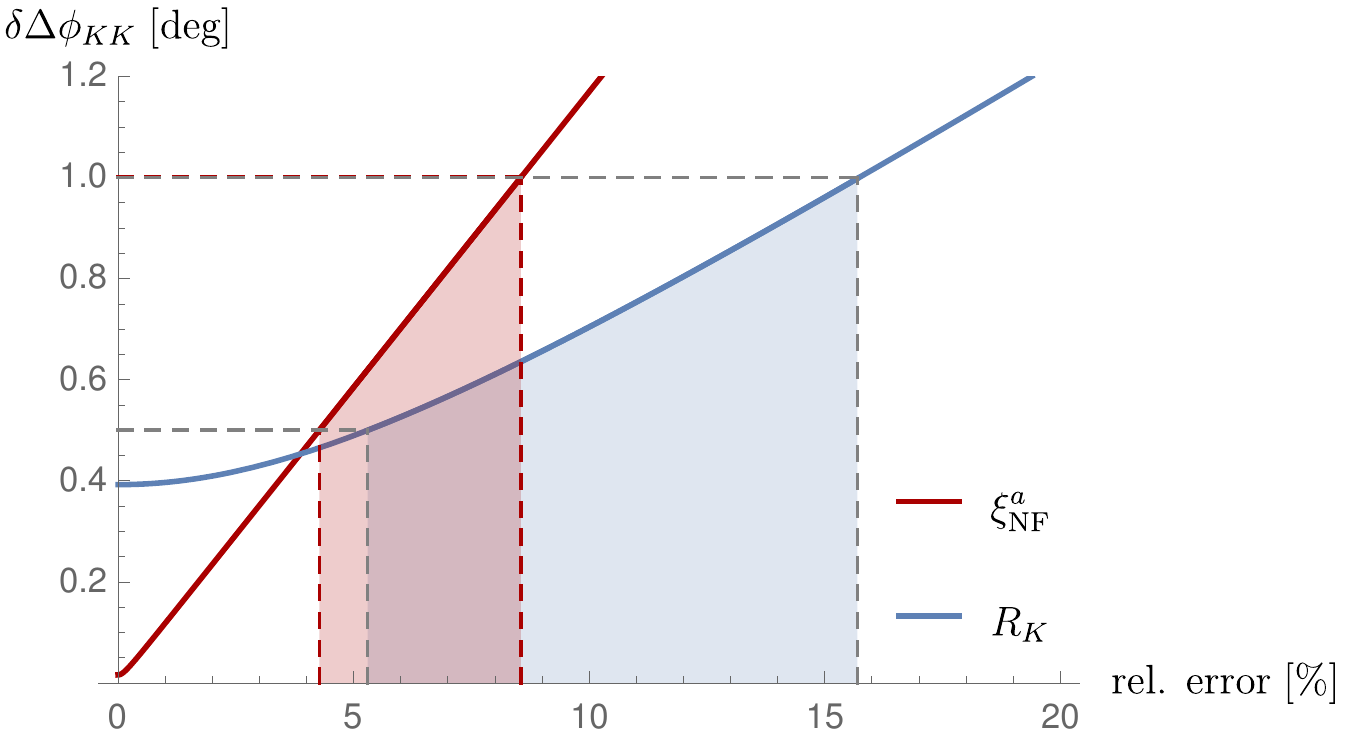}
	\caption{Illustration of the error on $\Delta\phi_{KK}$. For the dependence on 
	the $U$-spin-breaking parameter $\xi_{\rm NF}^a$ in (\ref{a-rel}) we assume a perfect 
	experimental situation, while a perfect theoretical situation is assumed for the dependence on 
	the relative error of $R_K$, assuming a precision for $R_\pi$ of $5\%$.}
	\label{fig:errorRK}
\end{figure}

Combining all these non-factorizable $U$-spin-breaking effects, we estimate the corresponding error 
of $\xi_{\rm NF}^a$ in (\ref{a-rel}) as $10\%$. Since $r_K\gg1$, as can be seen 
in (\ref{rK-def}), we obtain $d'\sim \epsilon\sqrt{r_K}\propto |a_{\rm NF}/a_{\rm NF}'|$. Since (\ref{DelPhiKK}) 
gives $\Delta\phi_{KK}\sim -10^\circ$ for $d'\sim0.6$, we conclude that this new method allows the 
determination of this phase with a theoretical precision at the $1^\circ$ level. Using experimental data, 
this error can be controlled in a more sophisticated way and even a regime of $0.5^\circ$ appears
achievable in the upgrade scenario \cite{BIG}. In Fig.~\ref{fig:errorRK},
we illustrate the error of $\Delta\phi_{KK}$. We observe that a precision of $0.5^\circ$ requires a 
measurement of both $R_\pi$ and $R_K$ with a relative precision of $5\%$ in an ideal
theoretical situation. A measurement of $R_K$ with a relative precision of $15\%$ would allow 
a precision of $1^\circ$, which would already be an impressive achievement.

The $U$-spin-breaking parameter $\xi$ in (\ref{U-break}), which limits the precision of the original 
method, can be written as
\begin{equation}
\xi=\xi_{\rm NF}^a
\left|\frac{T_{\rm fact}}{T'_{\rm fact}}\right|\biggl|\frac{P^{(ct)'}+PA^{(ct)'}}{P^{(ct)}+PA^{(ct)}}\biggr|.
\end{equation}
In contrast to $\xi_{\rm NF}^a$ in (\ref{a-rel}), $\xi$ involves penguin amplitudes with internal top and 
charm quarks, where also  ``charming penguins" enter \cite{charm-pen}. Since the leading 
$U$-spin-breaking corrections are associated with these contributions, the uncertainty is 
significantly larger than in the case of $\xi_{\rm NF}^a$, which governs the new strategy. 

Another key feature of this method is that we may actually determine both $\xi$ and $\Delta$
from the data, thereby allowing valuable insights into the $U$-spin symmetry at work. Assuming 
future determinations of $R_K$, $R_\pi$ and $\xi_{\rm NF}^a$ with $5\%$ precision, $\xi$ can be 
extracted with an uncertainty at the 0.07 level.  

The $B^0_s\rightarrow K^-\ell^+\nu_\ell$ decay has unfortunately not yet been measured. We strongly 
advocate analyses of this channel at Belle II and LHCb, aiming at a direct measurement of the ratio 
$R_\pi/ R_K$ which is required for our method. It is interesting to note that the ratio $f_s/f_d$ of 
the $B^0_{s,d}$ fragmentation functions, which is a key ingredient for measurements of branching 
ratios of $B^0_s$ mesons at hadron colliders \cite{FST-fs}, cancels in (\ref{RK}).

\section{Picture from Current Data}\label{sec:current-data}
In view of the lack of data for the determination of $R_K$ we consider $\BdtoKpi$, which arises
if we replace the spectator strange quark of $B^0_s\to K^-K^+$ by a down quark. This channel 
has only penguin and tree contributions. If we neglect the exchange and penguin annihilation topologies 
in $B^0_s\to K^-K^+$ and use the $SU(3)$ flavor symmetry, we get the following relation \cite{RF-07}:
\begin{equation}\label{d-tilde-rel}
d' e^{i\theta'}=\tilde d' e^{i\tilde\theta'},
\end{equation}
where $\tilde d'$, $\tilde\theta'$ are the $\BdtoKpi$ counterparts of $d'$, $\theta'$. As replacement for
$R_K$ we introduce
\begin{equation}
\tilde{R}_K \equiv \frac{ \Gamma(B^0_d\rightarrow \pi^-K^+)}{d\Gamma(B^0_d\rightarrow 
\pi^- \ell^+ \nu_\ell)/dq^2|_{q^2=m_K^2}} \ . 
\end{equation}
In the ratio $R_\pi/\tilde{R}_K$ the semileptonic decay rates cancel up to a small corrections due to the
different kinematical points. 

Using the current values $\gamma=(70\pm 7)^\circ$, $\phi_d = (43.2\pm1.8)^\circ$,  
$\mathcal{A}_{\textrm{CP}}^{\textrm{dir}}(B^0_d\rightarrow \pi^-K^+)=0.082 \pm 0.006$ 
and the CP asymmetries of $B^0_d\to\pi^-\pi^+$ in Table~\ref{table},  we obtain
\begin{equation}\label{d-theta-det}
d=0.58\pm0.16, \quad \theta=(151.4\pm7.6)^\circ
\end{equation}
\vspace*{-0.8truecm}
\begin{equation}\label{tilde-d-theta-det}
\tilde d'=0.51\pm 0.03, \quad \tilde\theta'=(157\pm2)^\circ,
\end{equation}
which yield 
\begin{equation}\label{xi-tilde}
\tilde\xi\equiv \frac{\tilde d'}{d}=0.88\pm0.20, \quad \tilde\Delta\equiv \tilde\theta'-\theta=(5.5\pm8.3)^\circ.
\end{equation}
Here the uncertainties correspond only to the input parameters. The agreement between (\ref{d-theta-det}) 
and (\ref{tilde-d-theta-det}) is remarkable, strongly disfavoring the anomalously large 
$U$-spin-breaking corrections of (50--100)\% considered in \cite{LHCb-BsKK-gam}. 

The current CP asymmetries of $B^0_s\to K^-K^+$ give $\phi_s^{\rm eff}=(-17.6\pm7.9)^\circ$.
Employing (\ref{d-tilde-rel}) results in
\begin{equation}\label{DelPhiKK-det}
\Delta\phi_{KK} = -(10.7\pm 0.6)^\circ \,.
\end{equation}
Consequently, we obtain 
\begin{equation}
\phi_s=\phi_s^{\rm eff}-\Delta\phi_{KK} = -(6.9 \pm 7.9)^\circ,
\end{equation}
where the uncertainty is fully dominated by experiment. This value of $\phi_s$ is in perfect 
agreement with (\ref{LHCb-res}). 

The analysis of the currently available data demonstrates impressively the power of the new 
strategy.

\section{Conclusions}
We have proposed a new strategy to extract the $B^0_s$--$\bar B^0_s$ mixing phase $\phi_s$ 
from the $B^0_s\to K^-K^+$, $B^0_d\to\pi^-\pi^+$ system. The novel ingredients are the semileptonic 
$B^0_s\rightarrow K^-\ell^+\nu_\ell$ and $B^0_d\rightarrow \pi^-\ell^+\nu_\ell$ decays, allowing us to
limit the application of the $U$-spin symmetry to theoretically favorable color-allowed tree 
amplitudes and robust quantities. This method provides a future determination of $\phi_s$ from the
CP violation in  $B^0_s\to K^-K^+$ with a theoretical precision as high as ${\cal O}(0.5^\circ)$, 
which matches the experimental prospects, and offers powerful tests of the $U$-spin symmetry. 
As there is currently no measurement of the $B^0_s\rightarrow K^-\ell^+\nu_\ell$ decay available, 
we used the $B_d^0\rightarrow \pi^-K^+$ mode to illustrate the new strategy and obtain a very 
promising picture from the current data. We strongly advocate experimental analyses of 
$B^0_s\rightarrow K^-\ell^+\nu_\ell$ and dedicated determinations of the $R_K$ and $R_\pi$ ratios. 
The comparison of $\phi_s$ extracted from the penguin-dominated $B^0_s\to K^-K^+$ decay with 
the SM prediction and alternative measurements may reveal new sources of CP violation. 
This  strategy offers exciting new opportunities for the era of Belle II and the LHCb upgrade.

\vspace*{0.5truecm}

%
%
%
%####################################################################
\section*{Acknowledgements} 
We would like to thank Kristof De Bruyn for very useful discussions. This work is supported by 
the Foundation for Fundamental Research on Matter (FOM) and by the 
Deutsche Forschungsgemeinschaft (DFG) within research unit FOR 1873 (QFET).
%####################################################################

%
%
%

%
%
%
\end{document}